\documentclass{JHEP}
\usepackage{epsfig}

\newcommand{\eref}[1]{(\ref{#1})}

\newcommand{\fref}[1]{Figure~\ref{#1}}
\newcommand{\cref}[1]{Chapter~\ref{#1}}
\newcommand{\beq}{\begin{equation}}
\newcommand{\eeq}{\end{equation}}
\newcommand{\ba}{\begin{array}}
\newcommand{\ea}{\end{array}}
\newcommand{\bcenter}{\begin{center}}
\newcommand{\ecenter}{\end{center}}

\def\IB{\relax\hbox{$\inbar\kern-.3em{\rm B}$}}
\def\IC{\relax\hbox{$\inbar\kern-.3em{\rm C}$}}
\def\ID{\relax\hbox{$\inbar\kern-.3em{\rm D}$}}
\def\IE{\relax\hbox{$\inbar\kern-.3em{\rm E}$}}
\def\IF{\relax\hbox{$\inbar\kern-.3em{\rm F}$}}
\def\IG{\relax\hbox{$\inbar\kern-.3em{\rm G}$}}
\def\IGa{\relax\hbox{${\rm I}\kern-.18em\Gamma$}}
\def\IH{\relax{\rm I\kern-.18em H}}
\def\IK{\relax{\rm I\kern-.18em K}}
\def\IL{\relax{\rm I\kern-.18em L}}
\def\IP{\relax{\rm I\kern-.18em P}}
\def\IR{\relax{\rm I\kern-.18em R}}
\def\IZ{\relax\ifmmode\mathchoice
{\hbox{\cmss Z\kern-.4em Z}}{\hbox{\cmss Z\kern-.4em Z}}
{\lower.9pt\hbox{\cmsss Z\kern-.4em Z}}
{\lower1.2pt\hbox{\cmsss Z\kern-.4em Z}}\else{\cmss Z\kern-.4em Z}\fi}
\def\II{\relax{\rm I\kern-.18em I}}

\def\sCC{{\kern 0.27em\vrule height1.45ex width0.03em depth0em
          \kern-0.30em\rm C}}
\def\C{{\mathchoice
  {\sCC}
  {\sCC}
  {\kern 0.225em \vrule height1.05ex width0.025em depth0em \kern-0.25em \rm C}
  {\kern 0.180em \vrule height0.78ex width0.02em depth0em \kern-0.2em \rm C}
        }}
\def\sHH{{\rm I\kern-.16em{}H}}
\def\H{{\mathchoice
  {\sHH}
  {\sHH}
  {\rm I\kern-.13em{}H}
  {\rm I\kern-.13em{}H} }}
\def\sNN{{\rm I\kern-.16em{}N}}
\def\N{{\mathchoice
  {\sNN}
  {\sNN}
  {\rm I\kern-.12em{}N}
  {\rm I\kern-.10em{}N} }}
\def\sPP{{\rm I\kern-.16em{}P}}
\def\P{{\mathchoice
  {\sPP}
  {\sPP}
  {\rm I\kern-.12em{}P}
  {\rm I\kern-.10em{}P} }}
\def\sQQ{{\kern 0.27em \vrule height1.45ex width0.03em depth0em
          \kern-0.30em \rm Q}}
\def\Q{{\mathchoice
        {\sQQ}
        {\sQQ}
  {\kern 0.225em \vrule height1.05ex width0.025em depth0em \kern-0.25em \rm Q}
  {\kern 0.180em \vrule height0.78ex width0.020em depth0em \kern-0.20em \rm Q}
        }}
\def\sRR{{\rm I\kern-0.16em{}R}}
\def\R{{\mathchoice
  {\sRR}
  {\sRR}
  {\rm I\kern-0.12em{}R}
  {\rm I\kern-0.10em{}R} }}
\def\sZZ{{\rm Z\kern-0.32em{}Z}}
\def\Z{{\mathchoice
  {\sZZ}
  {\sZZ} 
  {\rm Z\kern-0.3em{}Z}     
  {\rm Z\kern-0.25em{}Z} }}  
\def\ZZZ{{\rm Z\kern-0.24em{}Z}}
\def\sII{{\rm I\kern-0.16em{}I}}
\def\I{{\mathchoice
  {\sII}
  {\sII}
  {\rm I\kern-0.12em{}I}
  {\rm I\kern-0.10em{}I} }}

\def\inbar{\,\vrule height1.5ex width.4pt depth0pt}
\font\cmss=cmss10 \font\cmsss=cmss10 at 7pt

\def\smiley{\hbox{\large$\bigcirc$\hspace{-0.80em}\raise.2ex
\hbox{$\cdot\cdot$}\kern-.61em\lower.2ex\hbox{\scriptsize$\smile$}}\ }
\def\frowny{\hbox{\large$\bigcirc$\hspace{-0.80em}\raise.2ex
\hbox{$\cdot\cdot$}\kern-.635em\lower.2ex\hbox{\scriptsize$\frown$}}\ }

\def\I{{\rlap{1} \hskip 1.6pt \hbox{1}}}

\newcommand{\mat}[1]{\left( \matrix{#1} \right)}
\makeatletter
\let\hangafter\@hangfrom
\makeatother

\newtheorem{definition}{\sf DEFINITION}

\newtheorem{theorem}{\sf THEOREM}

\preprint{MIT-CTP-3070\\ \\ {\tt hep-th/}}

\title{Phase Structure of D-brane Gauge Theories and Toric Duality}
\author{Bo Feng, Amihay Hanany, and Yang-Hui He
\footnote{
Research supported in part
by the Reed Fund, the CTP and the LNS of MIT and the U.S. Department
of Energy under cooperative research agreement \# DE-FC02-94ER40818.
A. H. is also supported by an A. P. Sloan Foundation Fellowship,
a DOE OJI award. Y.-H. H. is also supported by the Presidential
Fellowship of MIT.}
\\
Center for Theoretical Physics,
\\ Massachusetts Institute of Technology,\\ Cambridge, MA 02139, USA.\\
\email{fengb, hanany, yhe@ctp.mit.edu}
}

\abstract{Harnessing the unimodular degree of freedom in the definition
of any toric diagram, we present a method of constructing inequivalent
gauge theories which are world-volume theories of D-branes probing
the same toric singularity. These theories are various {\em phases} in
partial resolution of Abelian orbifolds. As examples, two phases are
constructed for both the zeroth Hirzebruch and the second del Pezzo
surfaces. We show that such a phenomenon is a special case of
``Toric Duality'' proposed in 
\href{http://xxx.lanl.gov/abs/hep-th/0003085}{hep-th/0003085}.  
Furthermore, we investigate the general conditions that
distinguish these different gauge theories with the same 
(toric) moduli space.}
\begin{document}
\newpage
\section{Introduction}
The methods of toric geometry have been a crucial tool to the
understanding of many fundamental aspects of string theory on
Calabi-Yau manifolds (cf. e.g. \cite{Lecture}). 
In particular, the connexions between toric
singularities and
the manufacturing of various gauge theories as D-brane world-volume
theories have been intimate.

Such connexions have been motivated by a myriad of sources. 
As far back as
1993, Witten \cite{Witten} had shown, via the so-called gauged linear
sigma model, that the Fayet-Illiopoulos parametre $r$ in the D-term of
an ${\cal N}=2$ supersymmetric field theory with $U(1)$ gauge groups
can be tuned as an order-parametre which extrapolates between
the Landau-Ginzburg and Calabi-Yau phases of the theory, whereby
giving a precise viewpoint to the LG/CY-correspondence. What this
means in the context of Abelian gauge theories is that whereas for $r
\ll 0$, we have a Landau-Ginzberg description of the theory,
by taking $r \gg 0$, the space of classical vacua obtained from D- 
and F-flatness is described by a Calabi-Yau manifold, and in particular
a toric variety.

With the advent of D-brane technologies, vast amount of work has been
done to study the dynamics of world-volume theories on D-branes
probing various geometries. Notably, in \cite{Orb}, D-branes
have been used to probe Abelian singularities of the form
$\IC^2/\IZ_n$. Methods of studying the moduli space of the SUSY
theories describable by quiver diagrams have been developed by the
recognition of the Kronheimer-Nakajima ALE instanton construction,
especially the moment maps used therein \cite{Orb2}.

Much work followed \cite{KS,Morrison,LNV}. 
A key advance was made in \cite{DGM}, where,
exemplifying with Abelian $\IC^3$ orbifolds, a detailed method was
developed for capturing the various phases of the moduli space of the
quiver gauge theories as toric varieties.
In another vein, the huge factory built after the brane-setup approach
to gauge theories \cite{HW} has been continuing to elucidate the
T-dual picture of branes probing singularities (e.g. \cite{Han-Zaf,
Han-Ura, Han-He}). Brane setups for toric resolutions of $\IZ_2 \times
\IZ_2$, including the famous conifold, were addressed in
\cite{Uranga,Greene}. The general question of how to construct the quiver
gauge theory for an arbitrary toric singularity was still pertinent.
With the AdS/CFT correspondence emerging \cite{KS,Morrison},
the pressing need for the question arises again: 
given a toric singularity, how
does one determine the quiver gauge theory having the former as its
moduli space? 

The answer lies in ``Partial Resolution of Abelian Orbifolds'' and was
introduced and exemplified for the toric resolutions of the $\IZ_3 \times
\IZ_3$ orbifold \cite{DGM,Chris}. The method was subsequently presented in an
algorithmic and computationally feasible fashion in \cite{toric} and
was applied to a host of examples in \cite{Sarkar}.

One short-coming about the inverse procedure of going from the toric
data to the gauge theory data is that it is highly non-unique and in
general, unless one starts by partially resolving an orbifold
singularity, one would not be guaranteed with a physical world-volume
theory at all! Though the non-uniqueness was harnessed in \cite{toric}
to construct families of quiver gauge theories with the same toric
moduli space, a phenomenon which was dubbed ``toric duality,'' the
physicality issue remains to be fully tackled.

The purpose of this writing is to analyse toric duality within the
confinement of the canonical method of partial resolutions. Now we are
always guaranteed with a world-volume theory at the end and this
physicality is of great assurance to us. We find indeed that with the
restriction of physical theories, toric duality is still very much at
work and one can construct D-brane quiver theories that flow to the
same moduli space.

We begin in \S 2 with a seeming paradox which initially motivated our
work and which {\it ab initio} appeared to present a challenge to
the canonical method. In \S 3 we resolve the paradox by introducing the
well-known mathematical fact of toric isomorphisms. Then in \S 4, we
present a detailed analysis, painstakingly tracing through each step
of the inverse procedure to see how much degree of freedom one is
allowed as one proceeds with the algorithm. We consequently arrive at
a method of extracting torically dual theories which are all physical;
to these we refer as ``phases.''
As applications of these ideas in \S 5 we re-analyse the examples in
\cite{toric}, viz., the toric del Pezzo surfaces as well as the zeroth
Hirzebruch surface and find the various phases of the quiver gauge
theories with them as moduli spaces. Finally in \S 6 we end with
conclusions and future prospects.
\section{A Seeming Paradox}
In \cite{toric} we noticed the emergence of the phenomenon of ``Toric
Duality'' wherein the moduli space of vast numbers of gauge theories could
be parametrised by the {\em same} toric variety. Of course, as we
mentioned there, one needs to check extensively whether these theories
are all physical in the sense that they are world-volume theories of
some D-brane probing the toric singularity.

Here we shall discuss an issue of more immediate concern to the
physical probe theory. We recall that using the method of {\em partial
resolutions of Abelian orbifolds} \cite{toric,DGM,Chris,Uranga}, we could
always extract a canonical theory on the D-brane probing the
singularity of interest.

However, a discrepancy of results seems to have risen between
\cite{toric} and \cite{Morrison} on the precise world-volume theory of
a D-brane probe sitting on the zeroth Hirzebruch surface; let us
compare and contrast the two results here.
\begin{itemize}
\item Results from \cite{toric}: The matter contents of the theory are
given by (on the left we present the quiver diagram and on the right,
the incidence matrix that encodes the quiver):
$$
\begin{array}{lcr}
\mbox{
\epsfxsize=4cm
\epsfysize=4cm
\epsfbox{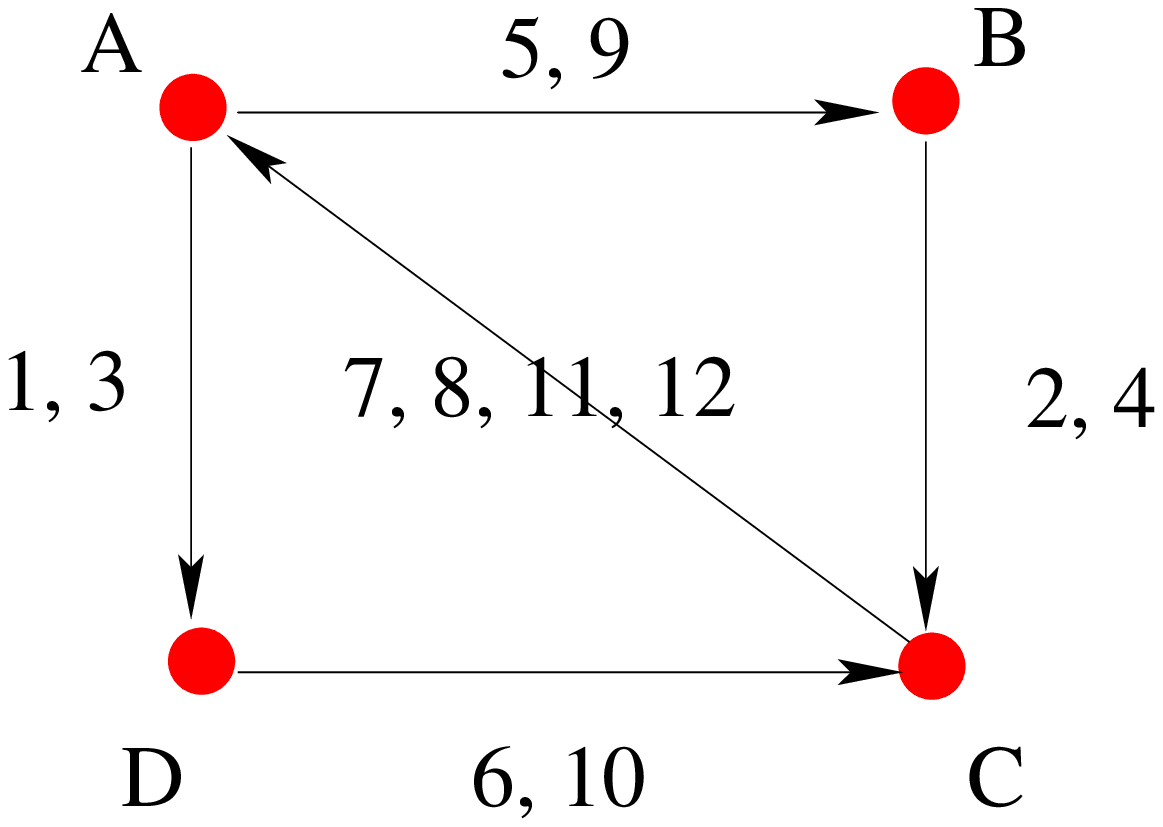}
}
\qquad \qquad
&&
d =
\mat{ \ba{c|cccccccccccc} 
	& X_1 & X_2 & X_3 & X_4 & X_5 & X_6 & X_7 & X_8 & X_9 & X_{10}
	& X_{11} & X_{12}\\ \hline
	A & -1 & 0 & -1 & 0 & -1 & 0 & 1 & 1 & -1 & 0 & 1 & 1 \\ 
	B & 0 & -1 & 0 & -1 & 1 & 0 & 0 & 0 & 1 & 0 & 0 & 0 \\
	C & 0 & 1 & 0 & 1 & 0 & 1 & -1 & -1 & 0 & 1 & -1 & -1 \\ 
	D & 1 & 0 & 1 & 0 & 0 & -1 & 0 & 0 & 0 & -1 & 0 & 0 \ea}
\end{array}
$$
and the superpotential is given by
\beq
W = X_{1}X_{8}X_{10}- X_{3}X_{7}X_{10}- X_{2}X_{8}X_{9}- X_{1}X_{6}X_{12}+\\ 
X_{3}X_{6}X_{11}+ X_{4}X_{7}X_{9}+ X_{2}X_{5}X_{12}- X_{4}X_{5}X_{11}.
\label{HirzeUs}
\eeq
\item Results from \cite{Morrison}: The matter contents of the theory are
given by (for $i=1,2$):
$$
\ba{lcr}
\epsfxsize=4cm
\epsfysize=4cm
\epsfbox{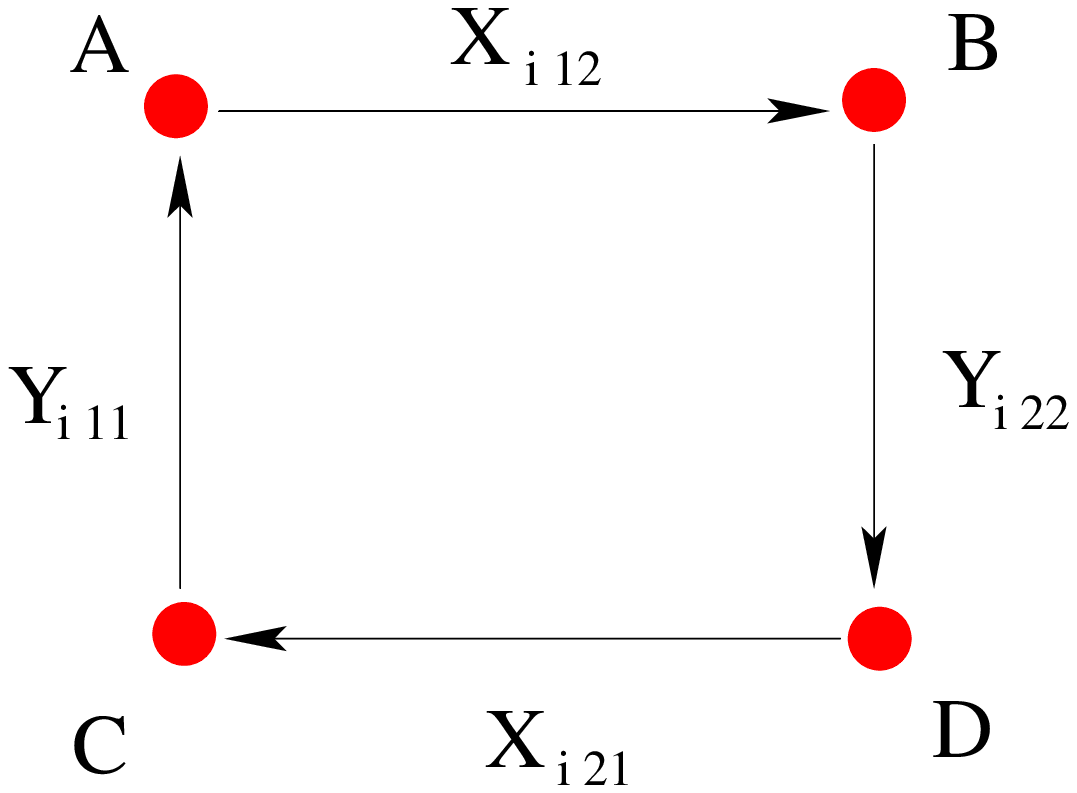}
\qquad \qquad
&&
d = 
\mat{\begin{array}{c|cccc}
         & X_{i~12}   & X_{i~21}    & Y_{i~11}  & Y_{i~22}   \\
\hline
A  &    -1     &     0    &   1     &    0  \\
B  &    1     &    0    &    0    &    -1   \\
C  &    0    &     1    &    -1    &    0   \\
D  &    0     &    -1    &   0    &    1   \\
\end{array}
}
\ea
$$
and the superpotential is given by 
\beq
W = \epsilon^{ij}
\epsilon^{kl}X_{i~12}Y_{k~22}X_{j~21}Y_{l~11}.
\label{HirzeMor}
\eeq
\end{itemize}

Indeed, even though both these theories have arisen from the canonical
partial resolutions technique and hence are world volume theories of a
brane probing a Hirzebruch singularity, we see clearly that they
differ vastly in both matter content and superpotential! Which is the
``correct'' physical theory?

In response to this seeming paradox, let us refer to \fref{f:Hirze}.
\EPSFIGURE[!ht]{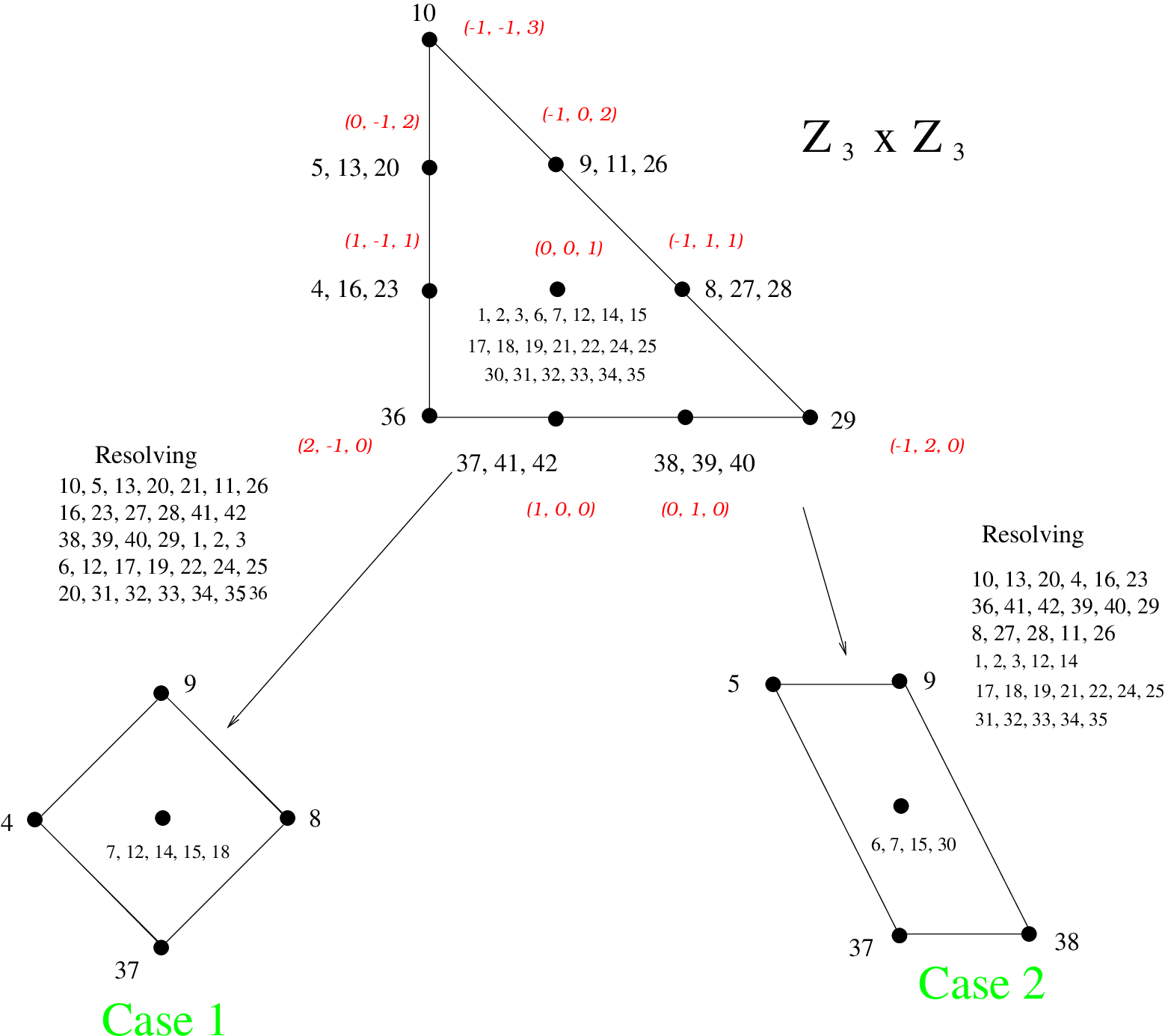,width=4.0in}
{
Two alternative resolutions of $\IC^2/\IZ_3 \times
\IZ_3$ to the Hirzebruch surface $F_0$: Case 1 from \cite{toric} and
Case 2 from \cite{Morrison}.
\label{f:Hirze}
}
Case 1 of course was what had been analysed in \cite{toric}
(q.v. ibid.) and presented in \eref{HirzeUs}; 
let us now consider case 2. Using the canonical
algorithm of \cite{Chris,toric}, we obtain the matter content (we have
labelled the fields and gauge groups with some foresight)
$$
d_{ia} = \left(
\ba{c|cccccccc}
	& X_1 & X'_1 & X'_2 & Y_1 & Y_2 & Y'_1 & Y_2 & Y'_2 \\ \hline
D & 0 & 1 & 1 & 0 & 0 & -1 & 0 & -1 \\
A & -1 & 0 & 0 & 1 & 1 & 0 & -1 & 0 \\
B & 1 & -1 & -1 & 0 & 0 & 0 & 1 & 0 \\
C & 0 & 0 & 0 & -1 & -1 & 1 & 0 & 1
\ea
\right)
$$
and the dual cone matrix
$$
K_{ij}^T =
\left(
\ba{c|cccccccc}
	& X_1 & X'_1 & X'_2 & Y_1 & Y_2 & Y'_1 & X_2 & Y'_2 \\ \hline
p_1 & 1 & 0 & 0 & 0 & 0 & 1 & 0 & 0 \\
p_2 & 0 & 1 & 0 & 1 & 0 & 0 & 0 & 0 \\ 
p_3 & 1 & 0 & 0 & 0 & 0 & 0 & 1 & 0 \\
p_4 & 0 & 1 & 1 & 0 & 0 & 0 & 0 & 0 \\
p_5 & 0 & 0 & 1 & 0 & 1 & 0 & 0 & 0 \\
p_6 & 0 & 0 & 0 & 0 & 0 & 1 & 0 & 1
\ea
\right)
$$
which translates to the F-term equations
$$
X_1 Y'_2 = p_1 p_3 p_6 = Y'_1 X_2; \quad
X'_1 Y_2 = p_2 p_4 p_5 = Y_1 X'_2.
$$
What we see of course, is that with the field redefinition
$X_i \leftrightarrow X_{i~12}, X'_i \leftrightarrow Y_{i~22},
Y_i \leftrightarrow Y_{i~11}$ and $Y'_i \leftrightarrow X_{i~21}$
for $i=1,2$, the above results are in exact agreement with the results
from \cite{Morrison} as presented in \eref{HirzeMor}.

This is actually of no surprise to us because upon closer inspection
of \fref{f:Hirze}, we see that the toric diagram for Cases 1 and 2
respectively has the coordinate points
$$
G1_t=
\left(
  \matrix{ -1 & 1 & 1 & 0 & -1 \cr 0 & 
     -1 & 0 & 0 & 1 \cr 2 & 1 & 0 & 1 & 1 \cr  } 
\right)
\qquad
G2_t=
\left(
  \matrix{ 0 & -1 & 1 & 0 & 0 \cr 
     -1 & 0 & 0 & 1 & 0 \cr 2 & 2 & 0 & 0 & 1 \cr  } 
\right).
$$
Now since the algebraic equation of the toric variety is given by
\cite{Fulton}
$$
V(G_t) = Spec_{Max}\left( \C[X_i^{G_t^\vee \cap \IZ^3}]\right),
$$
we have checked that, using a reduced Gr\"obner polynomial basis algorithm to
compute the variety \cite{Sturmfels}, the equations are
identical up to redefinition of variables.

Therefore we see that the two toric diagrams in Cases 1 and 2 of
\fref{f:Hirze} both describe the zeroth Hirzebruch surface as they
have the same equations (embedding into $\IC^9$). Yet due to the
particular choice of the diagram, we end up with strikingly different
gauge theories on the D-brane probe despite the identification of the
moduli space in the IR. This is indeed a curiously strong
version of ``toric duality.''

Bearing the above in mind, in this paper, 
we will analyse the degrees of freedom in the Inverse
Algorithm expounded upon in \cite{toric}, 
i.e., for a given toric singularity, how many different physical gauge
theories (phase structures), resulting from various partial
resolutions can one have for a D-brane probing such a singularity? 
To answer this question, first in
\S 2 we present the concept of toric isomorphism and give the
conditions for different toric data to correspond to the same toric 
variety. Then in \S 3 we follow the Forward Algorithm and give the 
freedom at each step from a given set of gauge theory data all the way
to the output of the toric data. 
Knowing these freedoms, we can identify the sources that may give rise
to different gauge theories in the Inverse Algorithm starting
from a prescribed toric data.
In section 4, we apply the above results and analyse the different
phases for the partial resolutions of the
$\IZ_3\times \IZ_3$ orbifold singularity, in particular,
we found that there are
two inequivalent phases of gauge theories respectively for the zeroth
Hirzebruch surface and the second del Pezzo surface.
Finally, in section 5, we give discussions for further investigation.
\section{Toric Isomorphisms}
Extending this observation to generic toric singularities, we expect
classes of inequivalent toric diagrams corresponding to the same
variety to give rise to inequivalent gauge theories on the D-brane
probing the said singularity. An immediate question is naturally
posed: ``is there a classification of these different theories and is
there a transformation among them?''

To answer this question we resort to the following result. 
Given $M$-lattice cones $\sigma$ and $\sigma'$, let the linear span of
$\sigma$ be lin$\sigma = \IR^n$ and that of $\sigma'$ be $\IR^m$. 
Now each cone gives rise to a semigroup which is the intersection of
the dual cone $\sigma^\vee$ with the dual lattice $M$, i.e., $S_\sigma
:= \sigma^\vee \cap M$ (likewise for $\sigma'$). Finally the toric
variety is given as the maximal spectrum of the polynomial ring of
$\IC$ adjoint the semigroup, i.e., $X_\sigma := Spec_{Max}
\left( \IC[S_\sigma]\right)$.
\begin{definition}We have these types of isomorphisms:
\begin{enumerate}
\item We call $\sigma$ and $\sigma'$ {\em cone isomorphic}, denoted
	$\sigma \cong_{cone} \sigma'$, if $n=m$ and there is
	a unimodular transformation $L:\IR^n \rightarrow \IR^n$ with
	$L(\sigma) = \sigma'$;
\item we call $S_\sigma$ and $S_{\sigma'}$ {\em monomial isomorphic},
	denoted $S_\sigma \cong_{mon} S_{\sigma'}$, if there
	exists mutually inverse monomial homomorphisms between the two
	semigroups.
\end{enumerate} 
\end{definition}
Thus equipped, we are endowed with the following
\begin{theorem}\label{iso}
(\cite{Ewald}, VI.2.11) The following conditions are equivalent:
$$
(a)~~\sigma \cong_{cone} \sigma' \Leftrightarrow
(b)~~S_\sigma \cong_{mon} S_{\sigma'} \Leftrightarrow
(c)~~X_\sigma \cong X_{\sigma'}
$$
\end{theorem}
What this theorem means for us is simply that, for the $n$-dimensional
toric variety, an $SL(n;\IZ)$ transformation\footnote{Strictly
	speaking, by unimodular we mean $GL(n;\IZ)$ matrices with
	determinant $\pm 1$; we shall denote these loosely by $SL(n;\IZ)$.
	}
on the original lattice
cone amounts to merely co\"ordinate transformations on the polynomial
ring and results in the same toric variety. This, is precisely what we
want: different toric diagrams giving the same variety.

The necessity and sufficiency of the condition in Theorem \ref{iso}
is important. Let us think of one example to
illustrate. Let a cone be defined by $(e_1,e_2)$, we know this
corresponds to $\IC^2$. Now if we apply the transformation
$$
(e_1,e_2) \left[ \begin{array}{cc} 2  & 0 \\ -1 & 1 \end{array}
\right]=(2e_1-e_2,e_2),
$$
which corresponds to the variety $xy=z^2$, i.e., $\IC^2/\IZ_2$, which of
course is not isomorphic to $\IC^2$. The reason for this is obvious:
the matrix we have chosen is certainly not unimodular.
\section{Freedom and Ambiguity in the Algorithm}
In this section, we wish to step back and address the issue in fuller
generality. Recall that the procedure of obtaining the moduli space
encoded as toric data once given the gauge
theory data in terms of product $U(1)$ gauge groups, D-terms from matter
contents and F-terms from the superpotential, has been well developed
\cite{Morrison,DGM}. Such was called the {\bf forward algorithm} in
\cite{toric}. On the other hand the {\bf reverse algorithm} of
obtaining the gauge theory data from the toric data has been discussed
extensively in \cite{Chris,toric}.

It was pointed in \cite{toric} that both the forward and reverse
algorithm are highly non-unique, a property which could actually be
harnessed to provide large classes of gauge theories having the same
IR moduli space.
In light of this so-named ``toric duality''
it would be instructive for us to investigate 
how much freedom do we have at each step in the algorithm.
We will call two data related by such a freedom {\em equivalent} to
each other. Thence further we could see how freedoms at every step
accumulate and appear in the final toric data. Modulo such
equivalences we believe that the data should be uniquely determinable.
\subsection{The Forward Algorithm}
We begin with the forward algorithm of extracting toric data from
gauge data. A brief review is at hand.
To specify the gauge theory, we require three pieces of information:
the number of $U(1)$ gauge fields, the charges of matter fields and the
superpotential.
The first two are summarised by the so-called charge matrix
$d_{li}$ where $l=1,2,...,L$ with $L$ the number of $U(1)$ gauge fields and
$i=1,2,...,I$ with $I$ the number of matter fields.
When using the forward algorithm to find the vacuum manifold (as a
toric variety), we need to solve the D-term and F-term flatness equations.
The D-terms are given by $d_{li}$ matrix while the 
F-terms are encoded in a matrix $K_{ij}$ with 
$i,1,2,...,I$ and $j=1,2,...,J$ where $J$ is the number of 
independent parameters needed to solve the F-terms. By gauge data then
we mean the matrices $d$ (also called the incidence matrix) and the
$K$ (essentially the dual cone); the forward algorithm takes these as input.
Subsequently we trace a flow-chart:
$$
\begin{array}{ccccccc}
\mbox{D-Terms} \rightarrow d     & \rightarrow   &\Delta & & & & \\
        &       &\downarrow     &       &       &       &       \\
\mbox{F-Terms} \rightarrow K    & \stackrel{V \cdot K^T =
        \Delta}{\rightarrow}
                & V      & & & & \\
\downarrow      &       & \downarrow    & & & & \\
T = {\rm Dual}(K)       & \stackrel{U \cdot T^T = {\rm
        Id}}{\rightarrow} & U & \rightarrow & VU & &\\
\downarrow      &       &       &       & \downarrow    & & \\
Q = [{\rm Ker}(T)]^T    &       & \longrightarrow       & & Q_t =
        \left( \begin{array}{c}
                                        Q \\ VU \end{array} \right) &
                                                \rightarrow & G_t =
        [{\rm Ker}(Q_t)]^T \\

\end{array}
$$
arriving at a final matrix $G_t$ whose columns are the vectors which
prescribe the nodes of the toric diagram.

What we wish to investigate below is how much procedural freedom we have at
each arrow so as to ascertain the non-trivial toric dual
theories. Hence, if $A_1$ is the matrix whither one arrives from a
certain arrow, then we would like to find the most general
transformation taking $A_1$ to another solution $A_2$ which would give
rise to an identical theory. It is to this transformation that we shall 
refer as ``freedom'' at the particular step.
\subsection*{Superpotential: the matrices $K$ and $T$}
The solution of F-term equations gives rise to a dual cone $K_1 = K_{ij}$
defined by $I$ vectors in $\IZ^J$.
Of course, we can choose different
parametres to solve the F-terms and arrive at another dual cone $K_2$. 
Then, $K_1$ and $K_2$, being integral cones, are equivalent if
they are unimodularly related, i.e., $ K_2^T= A \cdot K_1^T$ for
$A\in GL(J,\IZ)$ such that $\det(A)=\pm 1$.
Furthermore, the order of the $I$ vectors in
$\IZ^J$ clearly does not matter, so we can permute them by a matrix
$S_I$ in the symmetric group ${\cal S}_I$.
Thus far we have two freedoms, multiplication by $A$ and $S$:
\beq\label{K_tran}
K_2^T= A \cdot K_1^T \cdot S_I,
\eeq
and $K_{1,2}$ should give equivalent theories.

Now, from $K_{ij}$ we can find its dual matrix
$T_{j \alpha}$ (defining the cone $T$) where $\alpha=1,2,...,c$ and $c$ is the
number of vectors of the cone $T$ in $\IZ^J$, as constrained by
\beq \label{T_def}
K \cdot T \geq 0
\eeq
and such that $T$ also spans an integral cone.
Notice that finding dual cones, as given in a algorithm in
\cite{Fulton}, is actually unique up to
permutation of the defining vectors. 
Now considering the freedom of $K_{ij}$ as in
(\ref{K_tran}), let $T_2$ be the dual of $K_2$ and $T_1$ that of
$K_1$,  we have
$K_2 \cdot T_2= S_I^T \cdot  K_1 \cdot A^T \cdot T_2 \geq 0$,
which means that 
\beq
\label{T_tran}
T_1 = A^T \cdot T_2 \cdot S_c.
\eeq
Note that here $S_c$ is the permutation of the $c$ vectors of the cone $T$ in
and not that of the dual cone in \eref{K_tran}.
\subsection*{The Charge Matrix $Q$}
The next step is to find the charge matrix $Q_{k \alpha}$ where
$\alpha=1,2,...,c$ and $k=1,2,...,c-J$. This matrix is defined by
\beq
\label{Q_def}
T \cdot Q^T=0.
\eeq
In the same spirit as the above discussion, from \eref{T_tran} we have
$T_1 \cdot Q_1^T=A^T \cdot T_2 \cdot S_c \cdot Q_1^T=0$.
Because $A^T$ is a invertible matrix, this has a solution when and
only when $T_2 \cdot S_c \cdot Q_1^T=0$. Of course this is equivalent
to $T_2 \cdot S_c \cdot Q_1^T \cdot B_{kk'}=0$
for some invertible $(c-J)\times (c-J)$ matrix $B_{kk'}$. So the freedom for
matrix $Q$ is
\beq
\label{Q_tran}
Q_2^T=  S_c \cdot Q_1^T \cdot  B.
\eeq
We emphasize a difference from \eref{T_def}; there we required
both matrices $K$ and $T$ to be integer where here \eref{Q_def} does not
possess such a constraint. Thus the only condition for the matrix $B$ is
its invertibility.
\subsection*{Matter Content: the Matrices $d$, $\widetilde{V}$ and $U$}
Now we move onto the D-term and the integral $d_{li}$ matrix.
The D-term equations are $d \cdot |X|^2 = 0$ for matter fields $X$.
Obviously, any transformation on $d$ by an invertible matrix
$C_{L\times L}$ does not change the D-terms. Furthermore, any
permutation $S_I$ of the order the fields $X$, so long as it is
consistent with the $S_I$ in \eref{K_tran}, is also game.
In other words, we have the freedom:
\beq
\label{d_tran}
d_2 =C \cdot d_1 \cdot S_I.
\eeq
We recall that a matrix $V$ is then determined from $\Delta$, which is
$d$ with a row deleted due to the centre of mass degree of freedom.
However, to not to spoil the above freedom enjoyed by matrix $d$ in
\eref{d_tran}, we will make a slight
amendment and define the matrix $\widetilde{V}_{lj}$ by
\beq
\label{til_V_def}
\widetilde{V}\cdot K^T= d.
\eeq 
Therefore, whereas in \cite{DGM,toric} where $V\cdot K^T= \Delta$ 
was defined, we generalise $V$ to $\widetilde{V}$ by \eref{til_V_def}. 
One obvious way to obtain $\widetilde{V}$ from $V$ is to add one row such
that the sum of every column is zero. However, there is a caveat:
when there exists a vector $h$ such that
$$
h \cdot K^T=0,
$$
we have the freedom to add $h$ to any row of $\widetilde{V}$.
Thus finding the freedom of $\widetilde{V}_{lj}$ is a little more
involved. From \eref{K_tran} we have
$d_2= \widetilde{V}_2\cdot K_2^T=\widetilde{V}_2\cdot 
A \cdot K_1^T \cdot S_I$
and 
$d_2 =C \cdot d_1\cdot S_I =C \cdot \widetilde{V}_1\cdot K_1^T\cdot S_I$.
Because $S_I$ is an invertible square matrix, we have
$(\widetilde{V}_2\cdot A-C \cdot \widetilde{V}_1)\cdot K_1^T=0$,
which means 
$
\widetilde{V}_2\cdot A-C \cdot \widetilde{V}_1= CH_{K_1}
$
for a matrix $H$ constructed by having the aforementioned vectors $h$
as its columns.
When $K^T$ has maximal rank, $H$ is zero and this is in fact the more
frequently encountered situation.
However, when $K^T$ is not maximal rank, so as to give
non-trivial solutions of $h$,
we have that $\widetilde{V}_1 $ and $\widetilde{V}_2$ are equivalent if
\beq
\label{til_V_tran}
\widetilde{V}_2=C \cdot (\widetilde{V}_1+H_{K_1} ) \cdot A^{-1}.
\eeq

Moving on to the matrix $U_{j\alpha}$ defined by
\beq
\label{U_def}
U\cdot T^T= \II_{jj'},
\eeq
we have from \eref{T_tran}
$\II_{jj'}=U_1\cdot T_1^T=U_1\cdot S_c^T \cdot T_2^T \cdot A$,
whence $A^{-1}=U_1\cdot S_c^T \cdot T_2^T$ and 
$\II = A \cdot U_1\cdot S_c^T \cdot T_2^T$. This gives
$(A \cdot U_1\cdot S_c^T-U_2)\cdot T_2^T=0$
which has a solution
$A \cdot U_1\cdot S_c^T-U_2=H_{T_2}$
where $H_{T_2} \cdot T_2^T=0$ is precisely as defined in analogy of
the $H$ above. Therefore the freedom on $U$ is subsequently 
\beq
\label{U_tran}
U_2= A \cdot (U_1-H_{T_1})\cdot S_c^T,
\eeq
where $H_{T_1}=A^{-1} H_{T_2} (S_c^T)^{-1}$ and
$H_{T_1} \cdot T^T_1= (A^{-1} H_{T_2} (S_c^T)^{-1})(S_c^T \cdot T_2^T \cdot A)
=0$.
Finally using \eref{til_V_tran} and \eref{U_tran}, we have
\beq
\label{VU_tran}
(\widetilde{V}_2 \cdot U_2) =C \cdot (\widetilde{V}_1+H_{K_1})
 \cdot A^{-1} 
\cdot A \cdot (U_1-H_{T_1})\cdot S_c^T=
C \cdot (\widetilde{V}_1+H_{K_1})(U_1-H_{T_1}) \cdot S_c^T,
\eeq
determining the freedom of the relevant combination
$(\widetilde{V} \cdot U)$.

Let us pause for an important observation that in most cases 
$H_{K_1} = 0$, as we shall see in the examples later.
From \eref{Q_def}, which propounds the existence of a non-trivial
nullspace  for $T$, we see that one can indeed obtain
a non-trivial $H_{T_1}$ in terms of the combinations of the rows of
the charge matrix $Q$, whereby simplifying \eref{VU_tran} to
\beq
\label{VU_tran_1}
(\widetilde{V}_2 \cdot U_2)=C \cdot (\widetilde{V}_1 \cdot U_1+
H_{VU_1}) \cdot S_c^T,
\eeq
where every row of $H_{VU_1}$ is linear combination of rows of $Q_1$ and
the sum of its columns is zero.
\subsection*{Toric Data: the Matrices $Q_t$ and $G_t$}
At last we come to $\widetilde{Q}_t$, which is given by adjoining
$Q$ and $\widetilde{V}\cdot U$.  The freedom is of course, by
combining all of our results above,
\beq
\label{Qt_tran}
(\widetilde{Q}_t)_2 =\left( \begin{array}{c}  Q_2 \\ 
\widetilde{V}_2 \cdot U_2 \end{array} \right)
=\left( \begin{array}{c} B^T \cdot Q_1 \cdot S_c^T\\ 
C \cdot (\widetilde{V}_1 \cdot U_1+H_{VU_1}) \cdot S_c^T
\end{array} \right)
=\left( \begin{array}{c} B^T \cdot Q_1 \\ 
C \cdot (\widetilde{V}_1 \cdot U_1+H_{VU_1})
\end{array} \right) \cdot S_c^T
\eeq
Now $\widetilde{Q}_t$ determines the nodes of the toric diagram
$(G_t)_{p \alpha}$ ($p=1,2,..,(c-(L-1)-J)$ and $\alpha=1,2,...,c$) by 
\beq
\label{Gt_def}
Q_t \cdot G_t^T =0;
\eeq
The columns of $G_t$ then describes the toric diagram of the algebraic
variety for the vacuum moduli space and is the output of the algorithm.
From \eref{Gt_def} and  \eref{Qt_tran} we find that if 
$(\widetilde{Q}_t)_1 \cdot (G_t)_1^T=0$, i.e., 
$Q_1 \cdot (G_t)_1^T=0$ and $\widetilde{V}_1 \cdot U_1 \cdot
(G_t)_1^T=0,$ we automatically have the freedom $(\widetilde{Q}_t)_2 
\cdot (S_c^T)^{-1}\cdot (\widetilde{G}_t)_1^T=0$. This means that
at most we can have
\beq
\label{Gt_tran}
(G_t)_2^T=(S_c^T)^{-1}\cdot (G_t)_1^T \cdot D,
\eeq
where $D$ is a $GL(c-(L-1)-J,\IZ)$ matrix with $\det(D)=\pm 1$ which
is exactly the unimodular freedom for toric data as given by Theorem
\ref{iso}.

One immediate remark follows. From \eref{Gt_def} we obtain
the nullspace of $Q_t$ in $\IZ^c$. It seems that we can choose an
arbitrary basis so that $D$ is a $GL(c-(L-1)-J,\IZ)$ matrix with
the only condition that $\det(D)\neq 0$.
However, this is not stringent enough: in fact, when we find
cokernel $G_t$, we need to find the {\em integer basis} for the null
space, i.e., we need to find the basis such that any integer null vector can be 
decomposed into a linear combination of the columns of $G_t$. If we
insist upon such a choice, the only remaining freedom\footnote{
We would like to express our gratitude to M. Douglas for clarifying this point
to us.} is that $\det(D)=\pm 1$, viz, unimodularity.
\subsection{Freedom and Ambiguity in the Reverse Algorithm}
Having analysed the equivalence conditions in last subsection,
culminating in \eref{Qt_tran} and \eref{Gt_tran}, we now proceed in
the opposite direction and address the ambiguities in the reverse
algorithm.
\subsection*{The Toric Data: $G_t$}
We note that the $G_t$ matrix produced by the forward algorithm is
not minimal in the sense that certain columns are repeated, which
after deletion, constitute the toric diagram. 
Therefore, in our reverse algorithm, 
we shall first encounter such an ambiguity in deciding which columns
to repeat when constructing $G_t$ from the nodes of the toric
diagram. This so-called {\em repetition ambiguity} was discussed in
\cite{toric} and different choices of repetition may indeed
give rise to different gauge theories. It was pointed out ({\it
loc.~cit.}) that arbitrary repetition of the columns certainly does not
guarantee physicality. By physicality we mean that
the gauge theory arrived at the end of the day should be
{\em physical} in the sense of still being a D-brane world-volume theory.
What we shall focus here however, is the
inherent symmetry in the toric diagram, given by \eref{Gt_tran}, that
gives rise to the same theory. This is so that we could find truly
inequivalent {\em physical} gauge theories not related by
such a transformation as \eref{Gt_tran}.
\subsection*{The Charge Matrix: from $G_t$ to $Q_t$}
From \eref{Gt_def} we can solve for $Q_t$. However, for
a given $G_t$, in principle we can have two solutions $(Q_t)_1$ and
$(Q_t)_2$ related by
\beq \label{Qt_tran_rev}
(Q_t)_2 = P (Q_t)_1,
\eeq
where $P$ is a $p\times p$ matrix with $p$ the number of rows of
$Q_t$. Notice that the set of such transformations $P$ is much larger
than the counterpart in the forward algorithm given in
\eref{Qt_tran}. This is a second source of ambiguity in the reverse
algorithm. More explicitly,
we have the freedom to arbitrarily divide the $Q_t$ into two parts, viz.,
the D-term part $\widetilde{V}U$ and the F-term part $Q$.
Indeed one may find a matrix $P$ such that $(Q_t)_1$ and $(Q_t)_2$ satisfy
\eref{Qt_tran_rev} but not matrices $B$ and $C$ in order to satisfy 
\eref{Qt_tran}.
Hence different choices of $Q_t$ and different division therefrom into
D and F-term parts give rise to different gauge theories.
This is what we called {\em FD Ambiguity} in \cite{toric}. Again,
arbitrary division of the rows of $Q_t$ was pointed out to not to
ensure physicality. As with the discussion on the repetition ambiguity
above, what we shall pin down is the freedom due to the linear algebra
and not the choice of division.
\subsection*{The Dual Cone and Superpotential: from $Q$ to $K$}
The nullspace of $Q$ is the matrix $T$. The issue is the same as
discussed at the paragraph following \eref{Gt_tran} and one can
uniquely determine $T$ by imposing that its columns give an integral
span of the nullspace.
Going further from $T$ to its dual $K$, this is again a unique
procedure (while integrating back from $K$ to obtain the superpotential
is certainly not).
In summary then, these two steps give no sources for ambiguity.
\subsection*{The Matter Content: from $\widetilde{V}U$ to $d$ matrix}
The $d$ matrix can be directly calculated as \cite{toric}
\beq \label{d_def}
d=(\widetilde{V}U)\cdot T^T \cdot K^T.
\eeq
Substituting the freedoms in \eref{K_tran}, \eref{T_tran} and
\eref{VU_tran} we obtain
$$
\begin{array}{rcl}
d_2 & = & (\widetilde{V}_2 \cdot U_2) \cdot T_2^T \cdot K_2^T
=C \cdot [(\widetilde{V}_1 \cdot U_1)+H_{VU_1}] \cdot S_c^T
\cdot (S_c^T)^{-1} \cdot T_1^T \cdot A^{-1} \cdot A \cdot K_1^T \cdot S_I
\\ & & \\
& = & C \cdot (\widetilde{V}_1 \cdot U_1)\cdot T_1^T\cdot K_1^T\cdot S_I
+ C \cdot H_{VU_1} \cdot T_1^T\cdot K_1^T\cdot S_I
=C \cdot d_1 \cdot S_I,
\end{array}
$$
which is exactly formula (\ref{d_tran}). 
This  means that the matter matrices are equivalent up to a
transformation and there is no source for extra ambiguity.
\section{Application: Phases of $\IZ_3 \times \IZ_3$ Resolutions}
In \cite{toric} we developed an algorithmic outlook to the Inverse
Procedure and applied it to the construction of gauge theories on the
toric singularities which are partial resolutions of $\IZ_3 \times
\IZ_3$. The non-uniqueness of the method allowed one to obtain many
different gauge theories starting from the same toric variety, theories
to which we referred as being toric duals.
The non-uniqueness mainly comes from three
sources: (i) the repetition of the vectors in the toric data $G_t$
(Repetition Ambiguity), (ii) the different
choice of the null space basis of $Q_t$ and (iii) the different
divisions of the rows of $Q_t$ (F-D Ambiguity).
Many of the possible choices in the above will generate unphysical
gauge theories, i.e., not world-volume theories of D-brane probes. 
We have yet to catalogue the exact conditions which guarantee
physicality.

However, {\em Partial Resolution} of Abelian orbifolds, which stays
within subsectors of the latter theory, does indeed constrain the
theory to be physical. To these physical theories we shall refer as
{\bf phases} of the partial resolution. As discussed in \cite{toric}
any $k$-dimensional toric diagram can be embedded into $\IZ_n^{k-1}$
for sufficiently large $n$, one obvious starting point to
obtain different phases of a D-brane gauge theory is to try various
values of $n$. We leave some relevances of general $n$ to the Appendix.
However, because the algorithm of finding dual cones becomes
prohibitively computationally intensive even for $n \ge 4$, this
approach may not be immediately fruitful.

Yet armed with Theorem \ref{iso} we have an alternative. We can
certainly find all possible unimodular transformations of the given
toric diagram which still embeds into the same $\IZ_n^{k-1}$ and then
perform the inverse algorithm on these various {\it a fortiori}
equivalent toric data and observe what physical theories we obtain at
the end of the day. In our two examples in \S 1, we have essentially
done so; in those cases we found that two inequivalent gauge theory
data corresponded to two unimodularly equivalent toric data for the
examples of $\IZ_5$-orbifold and the zeroth Hirzebruch surface $F_0$.

The strategy lays itself before us. Let us illustrate with the same
examples as was analysed in \cite{toric}, namely the partial
resolutions of $\IC^3/(\IZ_3\times \IZ_3)$, i.e., $F_0$ and the toric del
Pezzo surfaces $dP_{0,1,2,3}$. We need to (i) find all $SL(3;\IZ)$
transformations of the toric diagram $G_t$ of these five singularities
that still remain as sub-diagrams of that of $\IZ_3\times \IZ_3$ and
then perform the inverse algorithm; therefrom, we must (ii) select
theories not related by any of the freedoms we have discussed above
and summarised in \eref{Qt_tran}.
\subsection{Unimodular Transformations within $\IZ_3\times \IZ_3$}
We first remind the
reader of the $G_t$ matrix of $\IZ_3\times \IZ_3$ given in
\fref{f:Hirze}, its columns are given by vectors:
$(0, 0, 1)$, $(1, -1, 1)$, $(0, -1, 2)$, $(-1, 1, 1)$, $(-1, 0, 2)$, 
$(-1, -1, 3)$, $(1, -1, 1)$, $(-1, 2, 0)$, $(1, 0, 0)$, $(0, 1, 0)$.
Step (i) of our above strategy can be immediately performed. Given the
toric data of one of the resolutions $G_t'$ with $x$ columns, 
we select $x$ from the above 10 columns of $G_t$ and check whether any
$SL(3;\IZ)$ transformation relates any permutation thereof
unimodularly to $G_t'$. We shall at the end find that there are
three different cases for $F_0$, five for $dP^0$, twelve for $dP_1$,
nine for $dP_2$ and only one for $dP_3$. The (unrepeated) $G_t$
matrices are as follows:
$$
\ba{|c|l|}
\hline
(F_0)_1 &  (0,0,1), (1,-1,1), (-1,1,1), (-1,0,2), (1,0,0) \\
(F_0)_2 & (0,0,1), (0,-1,2), (0,1,0), (-1,0,2), (1,0,0)\\
(F_0)_3 & (0,0,1), (1,-1,1), (-1,1,1), (0,-1,2), (0,1,0)\\
\hline
(dP_0)_1 & (0,0,1), (1,0,0),  (0,-1,2), (-1,1,1)\\
(dP_0)_2 & (0,0,1), (1,0,0), (-1,-1,3), (0,1,0)\\
(dP_0)_3 & (0,0,1), (-1,2,0), (1,-1,1), (0,-1,2)\\
(dP_0)_4 & (0,0,1), (0,1,0), (1,-1,1), (-1,0,2)\\
(dP_0)_5 & (0,0,1), (2,-1,0), (-1,1,1), (-1,0,2)\\
\hline
(dP_1)_1 & (1, 0, 0), (0, 1, 0), (-1, 1, 1), (0, -1, 2), (0, 0, 1)\\
(dP_1)_2 & (-1, -1, 3), (0, -1, 2), (1, 0, 0), (0, 1, 0), (0, 0, 1)\\
(dP_1)_3 & (0, -1, 2), (1, -1, 1), (1, 0, 0), (-1, 1, 1), (0, 0, 1)\\
(dP_1)_4 & (0, -1, 2), (1, -1, 1), (0, 1, 0), (-1, 2, 0), (0, 0, 1)\\
(dP_1)_5 & (0, -1, 2), (1, -1, 1), (0, 1, 0), (-1, 0, 2), (0, 0, 1)\\
(dP_1)_6 & (0, -1, 2), (1, -1, 1), (-1, 2, 0), (-1, 1, 1), (0, 0, 1)\\
(dP_1)_7 & (0, -1, 2), (1, 0, 0), (-1, 1, 1), (-1, 0, 2), (0, 0, 1)\\
(dP_1)_8 & (1, -1, 1), (2, -1, 0), (-1, 1, 1), (-1, 0, 2), (0, 0, 1)\\
(dP_1)_9 & (1, -1, 1), (1, 0, 0), (0, 1, 0), (-1, 0, 2), (0, 0, 1)\\
(dP_1)_{10} & (1, -1, 1), (0, 1, 0), (-1, 1, 1), (-1, 0, 2), (0, 0, 1)\\
(dP_1)_{11} & (2, -1, 0), (1, 0, 0), (-1, 1, 1), (-1, 0, 2), (0, 0, 1)\\
(dP_1)_{12} & (-1, -1, 3), (1, 0, 0), (0, 1, 0), (-1, 0, 2), (0, 0,1)\\
\hline
(dP_2)_1 & (2, -1, 0), (1, -1, 1), (-1, 0, 2), (-1, 1, 1), (1, 0, 0), (0, 0, 1)\\
(dP_2)_2 & (-1, -1, 3), (0, -1, 2), (1, 0, 0), (0, 1, 0), (-1, 0, 2), (0, 0, 1)\\
(dP_2)_3 & (0, -1, 2), (1, -1, 1), (1, 0, 0), (0, 1, 0), (-1, 1, 1), (0, 0, 1)\\
(dP_2)_4 & (0, -1, 2), (1, -1, 1), (1, 0, 0), (0, 1, 0), (-1, 0, 2), (0, 0, 1)\\
(dP_2)_5 & (0, -1, 2), (1, -1, 1), (1, 0, 0), (-1, 1, 1), (-1, 0, 2), (0, 0, 1)\\
(dP_2)_6 & (0, -1, 2), (1, -1, 1), (0, 1, 0), (-1, 2, 0), (-1, 1, 1), (0, 0, 1)\\
(dP_2)_7 & (0, -1, 2), (1, -1, 1), (0, 1, 0), (-1, 1, 1), (-1, 0, 2), (0, 0, 1)\\
(dP_2)_8 & (0, -1, 2), (1, 0, 0), (0, 1, 0), (-1, 1, 1), (-1, 0, 2), (0, 0, 1)\\
(dP_2)_9 & (1, -1, 1), (1, 0, 0), (0, 1, 0), (-1, 1, 1), (-1, 0, 2), (0, 0, 1)\\
\hline
dP_3 & (0, -1, 2), (1, -1, 1), (1, 0, 0), (0, 1, 0), 
	(-1, 1, 1), (-1, 0, 2), (0, 0, 1)\\
\hline
\ea
$$
The reader is referred to \fref{f:F0} to \fref{f:dP3} 
for the toric diagrams of the data
above. The vigilant would of course recognise $(F_0)_1$ to be Case 1
and $(F_0)_2$ as Case 2 of \fref{f:Hirze} as discussed in \S 2 and
furthermore $(dP_{0,1,2,3})_1$ to be the cases addressed in \cite{toric}.
\vspace{1in}
\subsection{Phases of Theories}
The Inverse Algorithm can then be readily applied to the above toric
data; of the various unimodularly equivalent toric diagrams of the del Pezzo
surfaces and the zeroth Hirzebruch, the details of which fields remain
massless at each node (in the
notation of \cite{toric}) are also presented in those figures
immediately referred to above.
\EPSFIGURE[!ht]{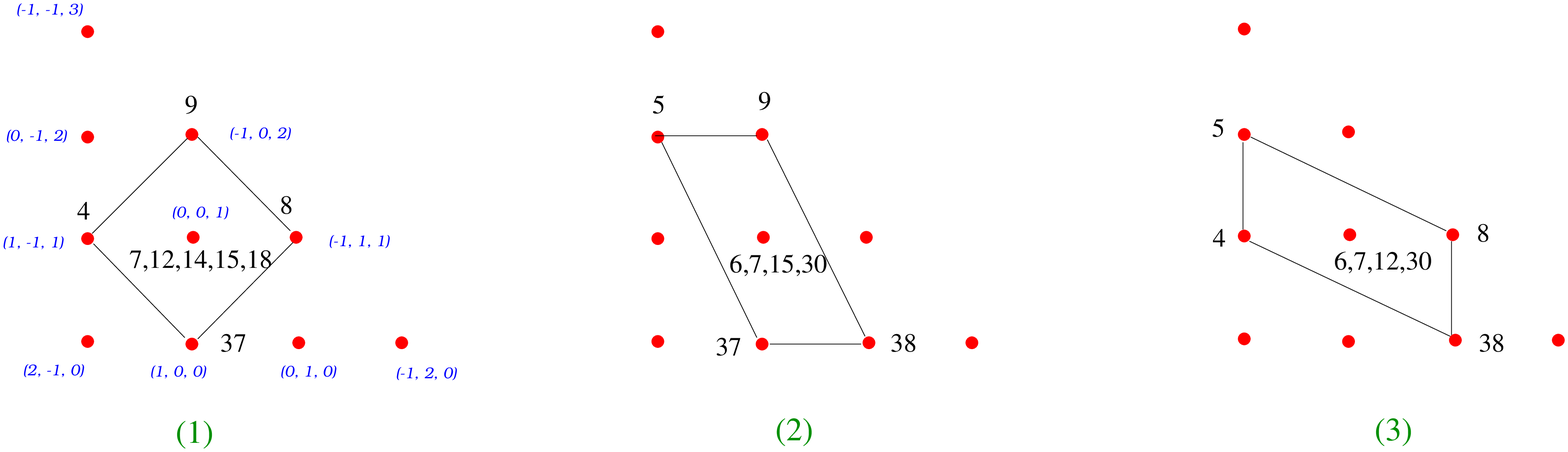,width=6in}
{The 3 equivalent representations of the toric diagram of the zeroth
Hirzebruch surface as a resolution of $\IZ_3\times \IZ_3$. We see that
(2) and (3) are related by a reflection about the $45^o$ line (a symmetry
inherent in the parent $\IZ_3 \times \IZ_3$ theory) and we
have the two giving equivalent gauge theories as expected.
\label{f:F0}
}
\EPSFIGURE[!ht]{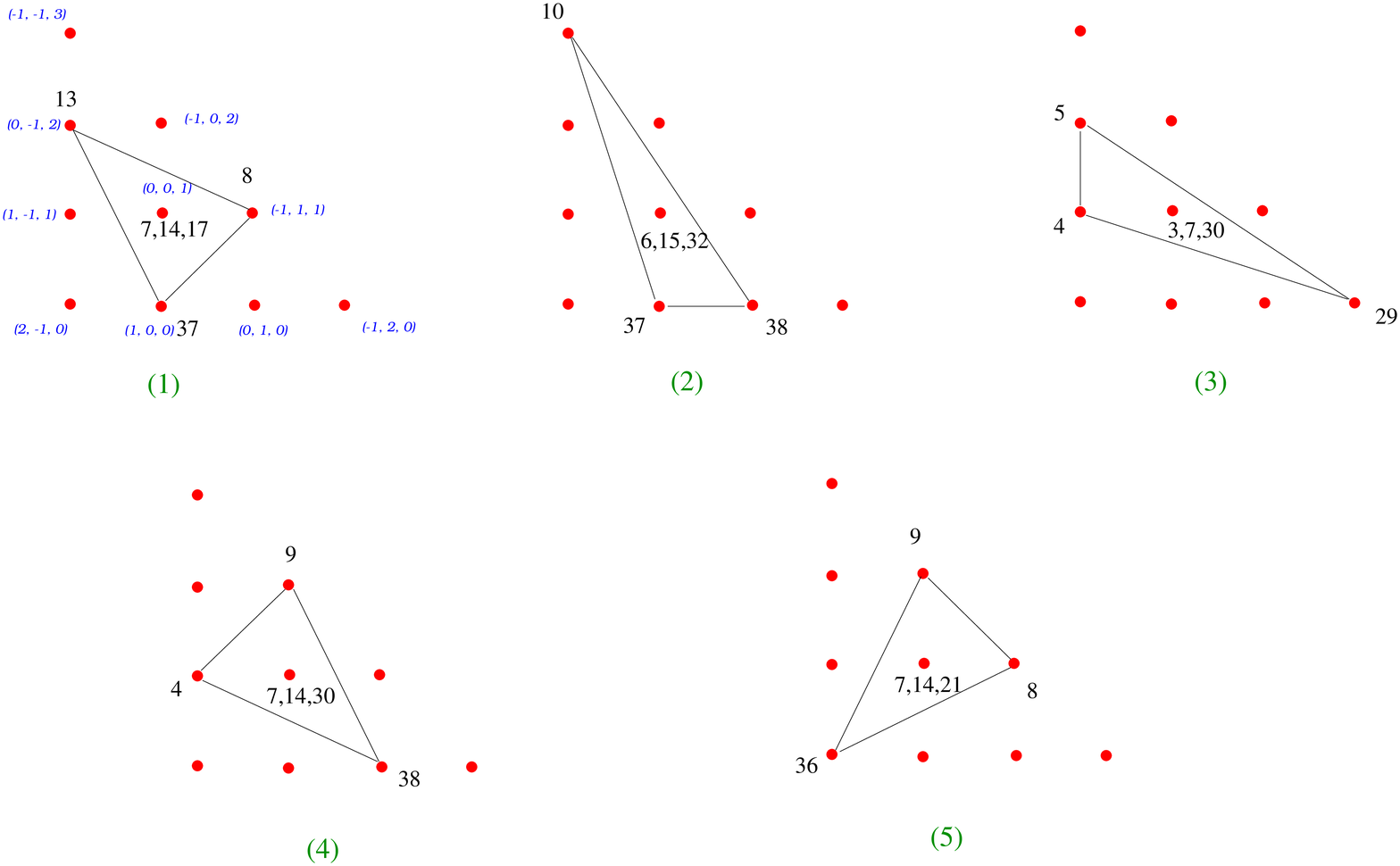,width=5.8in}
{The 5 equivalent representations of the toric diagram of the zeroth
del Pezzo surface as a resolution of $\IZ_3\times \IZ_3$. Again (1)
and (4) (respectively (2) and (3)) are related by the $45^o$
reflection, and hence give equivalent theories. In fact further
analysis shows that all 5 are equivalent.
\label{f:dP0}
}
\EPSFIGURE[!ht]{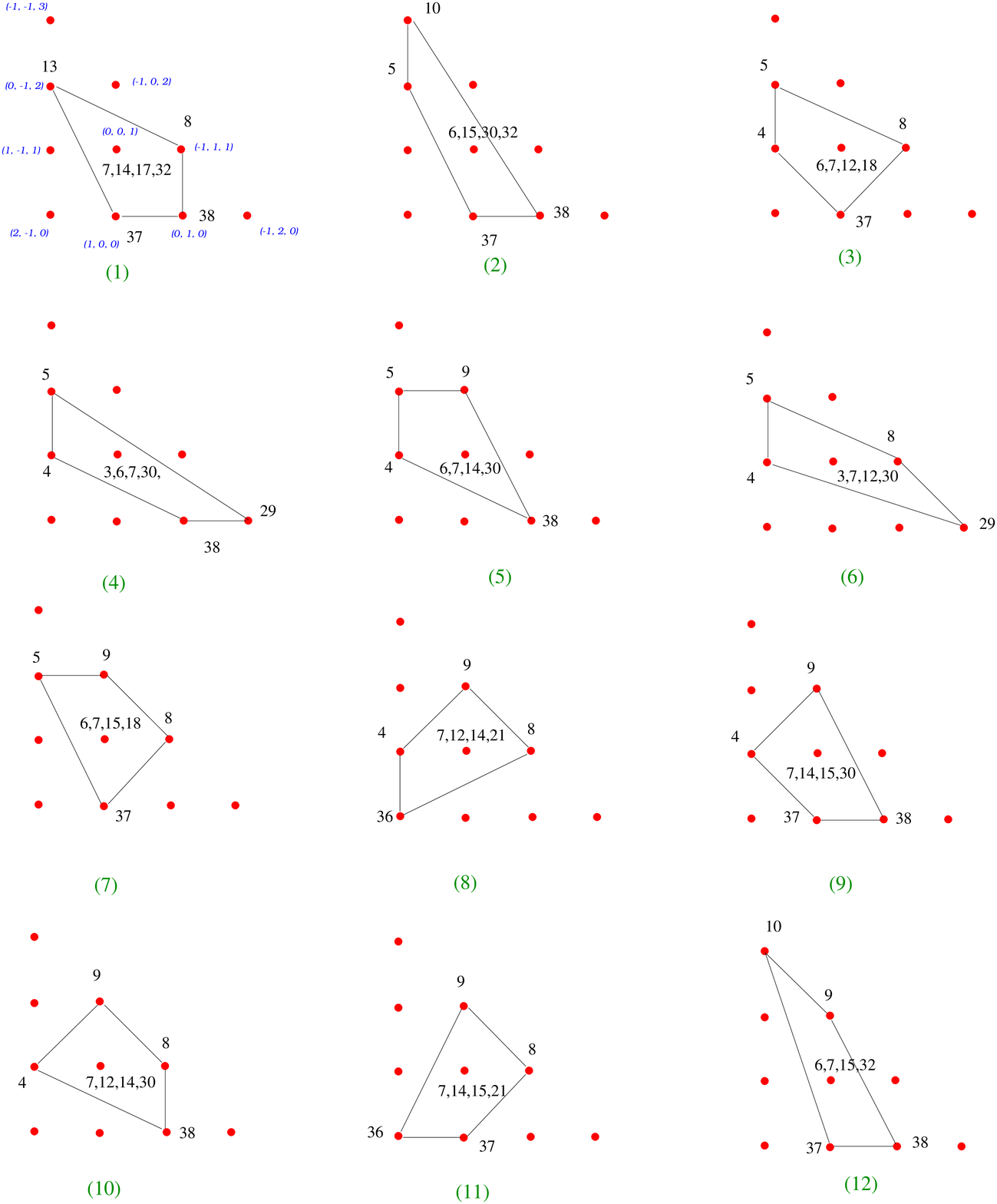,width=6in}
{The 12 equivalent representations of the toric diagram of the first
del Pezzo surface as a resolution of $\IZ_3\times \IZ_3$. The pairs
(1,5); (2,4); (3,9); (6,12); (7,10) and (8,11) are each reflected by
the $45^o$ line and give mutually equivalent gauge theories
indeed. Further analysis shows that all 12 are equivalent.
\label{f:dP1}
}
\EPSFIGURE[!ht]{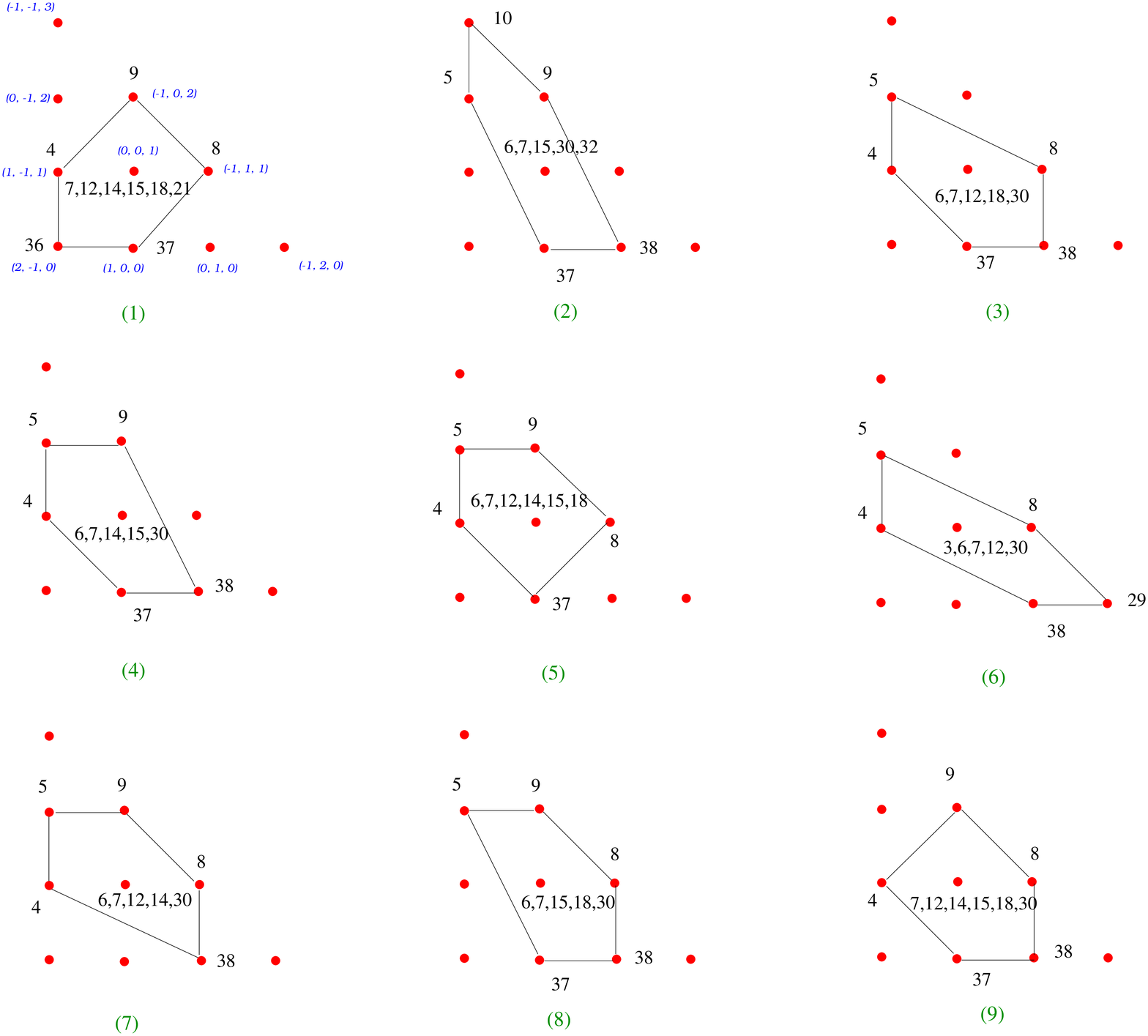,width=6in}
{The 9 equivalent representations of the toric diagram of the second
del Pezzo surface as a resolution of $\IZ_3\times \IZ_3$. The pairs
(2,6); (3,4); (5,9) and (7,8) are related by $45^o$ reflection while
(1) is self-reflexive and are hence give pairwise equivalent
theories. Further analysis shows that there are two phases given
respectively by (1,5,9) and (2,3,4,6,7,8).
\label{f:dP2}
}
\EPSFIGURE[!ht]{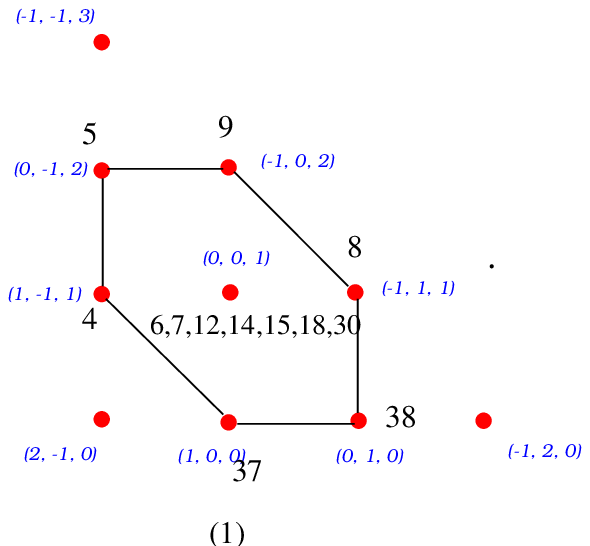,width=5in}
{The unique representations of the toric diagram of the third
del Pezzo surface as a resolution of $\IZ_3\times \IZ_3$.
\label{f:dP3}
}

Subsequently, we arrive at a number of D-brane gauge theories;
among them, all five cases for $dP^0$ are equivalent (which is in
complete consistency with the fact that $dP^0$ is simply
$\IC^3/\IZ_3$ and there is only one
nontrivial theory for this orbifold, corresponding to the decomposition 
${\bf 3}\rightarrow 1 + 1 + 1$). 
For $dP_1$, all twelve cases give back
to same gauge theory (q.v. Figure 5 of \cite{toric}). 
For $F_0$, the three cases give two inequivalent
gauge theories as given in \S 2. Finally
for $dP_2$, the
nine cases again give two different theories. For reference we
tabulate the D-term matrix $d$ and F-term matrix $K^T$ below. If more
than 1 theory are equivalent, then we select one representative from
the list, the matrices for the rest are given by transformations
\eref{K_tran} and \eref{d_tran}.
\newpage
\[
\hspace{-0.8in}
\ba{|c|c|c|}
\hline
\mbox{Singularity} & \mbox{Matter Content }d & \mbox{Superpotential}\\
\hline
(F_0)_1 & {\tiny \ba{c|cccccccccccc} 
	& X_1 & X_2 & X_3 & X_4 & X_5 & X_6 & X_7 & X_8 & X_9 & X_{10}
	& X_{11} & X_{12}\\ \hline
	A & -1 & 0 & -1 & 0 & -1 & 0 & 1 & 1 & -1 & 0 & 1 & 1 \\ 
	B & 0 & -1 & 0 & -1 & 1 & 0 & 0 & 0 & 1 & 0 & 0 & 0 \\
	C & 0 & 1 & 0 & 1 & 0 & 1 & -1 & -1 & 0 & 1 & -1 & -1 \\ 
	D & 1 & 0 & 1 & 0 & 0 & -1 & 0 & 0 & 0 & -1 & 0 & 0 \ea } 
	&
{\footnotesize
\ba{r}
X_{1}X_{8}X_{10}- X_{3}X_{7}X_{10}- X_{2}X_{8}X_{9}- X_{1}X_{6}X_{12}+\\ 
X_{3}X_{6}X_{11}+ X_{4}X_{7}X_{9}+ X_{2}X_{5}X_{12}- X_{4}X_{5}X_{11}
\ea
}
\\ \hline
(F_0)_{2,3} & {\tiny \ba{c|cccccccc}
	& X_{112} & Y_{122} & Y_{222} & Y_{111} & Y_{211} & X_{121} &
X_{212} & X_{221} \\ \hline
	A & -1 & 0 & 0 & 1 & 1 & 0 & -1 & 0 \\ 
	B & 1 & -1 & -1 & 0 & 0 & 0 & 1 & 0 \\ 
	C & 0 & 0 & 0 & -1 & -1 & 1 & 0 & 1 \\
	D & 0 & 1 & 1 & 0 & 0 & -1 & 0 & -1 \ea }
& 
{\tiny
\epsilon^{ij} \epsilon^{kl}X_{i~12}Y_{k~22}X_{j~21}Y_{l~11}
}
\\ \hline \hline
(dP_0)_{1,2,3,4,5} & {\tiny \ba{c|ccccccccc}
	& X_1 & X_2 & X_3 & X_4 & X_5 & X_6 & X_7 & X_8 & X_9 \\ \hline
	A & -1 & 0 & -1 & 0 & -1 & 0 & 1 & 1 & 1 \cr 
	B & 0 & 1 & 0 & 1 & 0 & 1 & -1 & -1 & -1 \cr 
	C & 1 & -1 & 1 & -1 & 1 & -1 & 0 & 0 & 0 \ea } 
&
{\footnotesize
\ba{r}
X_{1} X_{4} X_{9} - X_{4} X_{5} X_{7} - X_{2} X_{3} X_{9} - \\
X_{1} X_{6} X_{8} + X_{2} X_{5} X_{8} + X_{3} X_{6} X_{7}
\ea
}
\\ \hline \hline
(dP_1)_{1,2,...,12} &
	{\tiny	\ba{c|cccccccccc}
	& X_1 & X_2 & X_3 & X_4 & X_5 & X_6 & X_7 & X_8 & X_9 & X_{10}
	\\ \hline 
	A & -1 & 0 & 0 & -1 & 0 & 0 & 0 & 1 & 0 & 1 \cr
	B & 1 & -1 & 0 & 0 & 0 & -1 & 0 & 0 & 1 & 0 \cr
	C & 0 & 0 & 1 & 0 & 1 & 0 & 1 & -1 & -1 & -1 \cr 
	D & 0 & 1 & -1 & 1 & -1 & 1 & -1 & 0 & 0 & 0 \ea }
& 
{\footnotesize
\ba{r}
X_{2} X_{7} X_{9} - X_{3} X_{6} X_{9} - X_{4} X_{8} 
        X_{7} - X_{1} X_{2} X_{5} X_{10} \\
        + X_{3} X_{4} X_{10} + X_{1} X_{5} X_{6} X_{8}
\ea}
\\ \hline \hline
(dP_2)_{1,5,9} &
	{\tiny \ba{c|ccccccccccccc}
	& X_1 & X_2 & X_3 & X_4 & X_5 & X_6 & X_7 & X_8 & X_9 & X_{10}
	& X_{11} & X_{12} & X_{13} \\ \hline
	A & -1 & 0 & 0 & -1 & 0 & -1 & 0 & 1 & 0 & 0 & 0 & 1 & 1 \cr
	B & 0 & 0 & -1 & 0 & -1 & 1 & 0 & 0 & 0 & 1 & 0 & 0 & 0 \cr
	C & 0 & 0 & 1 & 0 & 1 & 0 & 1 & -1 & -1 & 0 & 1 & -1 & -1 \cr 
	D & 1 & -1 & 0 & 0 & 0 & 0 & 0 & 0 & 1 & -1 & 0 & 0 & 0 \cr 
	E & 0 & 1 & 0 & 1 & 0 & 0 & -1 & 0 & 0 & 0 & -1 & 0 & 0 \ea}
&
{\footnotesize
\ba{r}
X_{2} X_{9} X_{11} - X_{9} X_{3} X_{10} - X_{4} X_{8} X_{11} -
X_{1} X_{2} X_{7} X_{13} + X_{13} X_{3} X_{6} \\
- X_{5} X_{12} X_{6}+
X_{1} X_{5} X_{8} X_{10} + X_{4} X_{7} X_{12}
\ea
}
\\ \hline
(dP_2)_{2,3,4,6,7,8} &
	{\tiny \ba{c|cccccccccccc}
	& X_1 & X_2 & X_3 & X_4 & X_5 & X_6 & X_7 & X_8 & X_9 & X_{10}
	& X_{11} \\ \hline
	A & -1 & 0 & -1 & 0 & 0 & 0 & 1 & 0 & 0 & 0 & 1 \cr
	B & 1 & -1 & 0 & 0 & -1 & 0 & 0 & 0 & 1 & 0 & 0 \cr
	C & 0 & 0 & 1 & -1 & 0 & 1 & 0 & 0 & -1 & 0 & 0 \cr
	D & 0 & 0 & 0 & 0 & 0 & -1 & -1 & 1 & 0 & 1 & 0 \cr
	E & 0 & 1 & 0 & 1 & 1 & 0 & 0 & -1 & 0 & -1 & -1 \ea}
&
{\footnotesize
\ba{r}
X_5 X_8 X_6 X_9 + X_1 X_2 X_{10} X_7 + X_{11} X_3 X_4 \\
- X_4 X_{10} X_6 - X_2 X_8 X_7 X_3 X_9 - X_{11} X_1 X_5
\ea
}
\\ \hline \hline
(dP_3)_1 & {\tiny \ba{c|cccccccccccccc}
	& X_1 & X_2 & X_3 & X_4 & X_5 & X_6 & X_7 & X_8 & X_9 & X_{10}
	& X_{11}& X_{12}& X_{13}& X_{14} \\ \hline
	A & -1 & 0 & 0 & 0 & 1 & 0 & 0 & 1 & -1 & 0 & 0 & 1 & -1 & 0 \cr 
	B & 0 & 0 & -1 & 1 & 0 & -1 & 0 & 0 & 0 & 0 & 0 & 0 & 1 & 0\cr 
	C & 1 & -1 & 0 & -1 & 0 & 0 & 0 & 0 & 0 & 0 & 0 & 0 & 0 & 1\cr 
	D & 0 & 0 & 1 & 0 & 0 & 0 & 0 & -1 & 0 & -1 & 1 & 0 & 0 & 0\cr 
	E & 0 & 0 & 0 & 0 & -1 & 1 & 1 & 0 & 0 & 1 & 0 & -1 & 0 & -1\cr 
	F & 0 & 1 & 0 & 0 & 0 & 0 & -1 & 0 & 1 & 0 & -1 & 0 & 0 & 0 \ea}
&
{\footnotesize
\ba{r}
X_{3} X_{8} X_{13} - X_{8} X_{9} X_{11} - X_{5} X_{6} X_{13} - 
X_{1} X_{3} X_{4} X_{10} X_{12} \\
+ X_{7} X_{9} X_{12} + X_{1} X_{2} X_{5} X_{10} X_{11} + 
X_{4} X_{6} X_{14} - X_{2} X_{7} X_{14}
\ea
}
\\ \hline
\ea
\]
The matter content for these above theories are represented as quiver
diagrams in \fref{f:quiver} (multi-valence arrows are labelled with a
number) and the superpotentials, in the table below.
\EPSFIGURE[h]{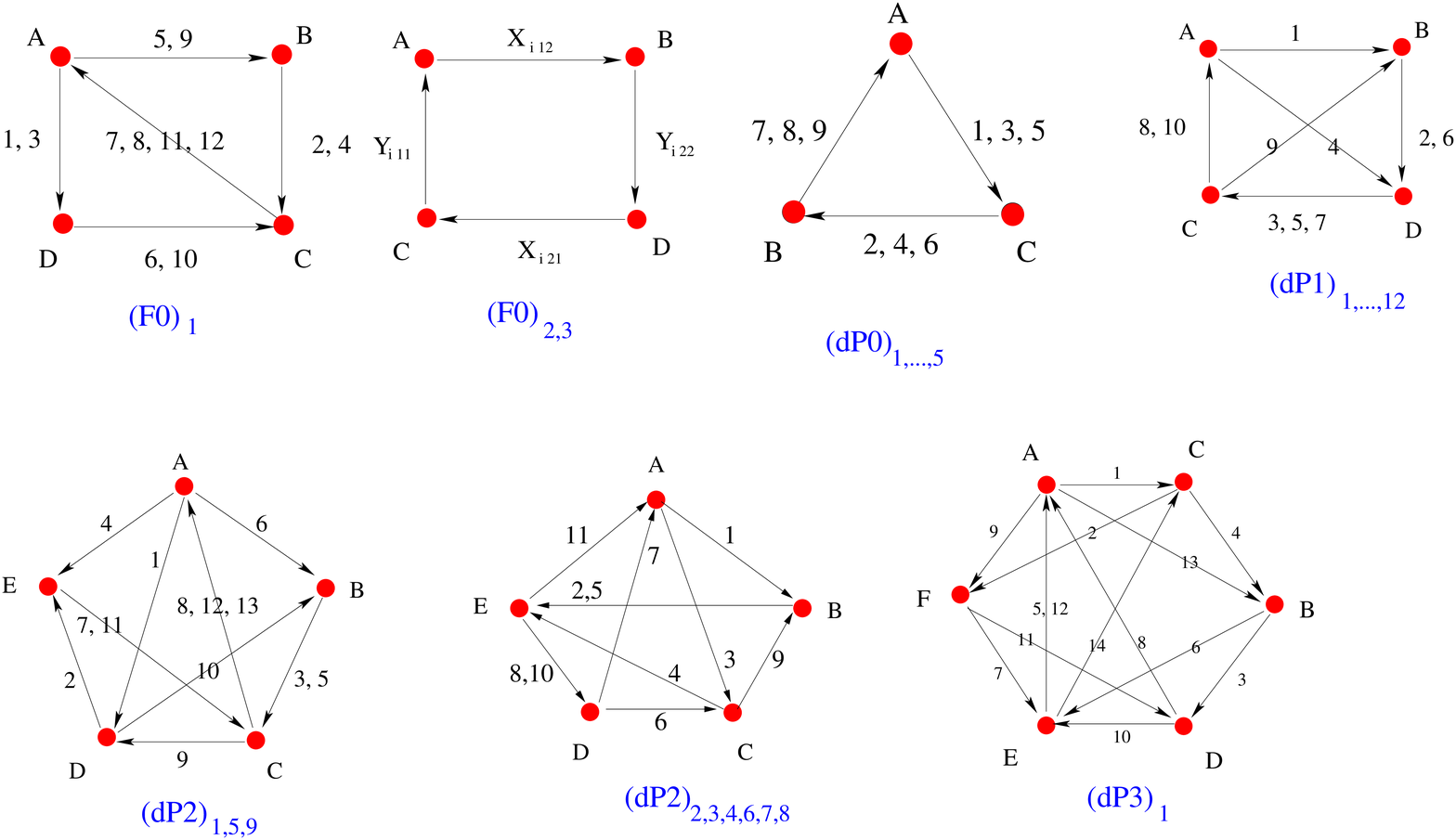,width=6.5in}
{The quiver diagrams for the various phases of the gauge theory for
the del Pezzo surfaces and the zeroth Hirzebruch surface.
\label{f:quiver}
}

In all of the above discussions, we have restricted ourselves to the
cases of $U(1)$ gauge groups, i.e., with only a single brane probe;
this is because such is the only case to which the toric technique can
be applied.
However, after we obtain the matter contents and superpotential for
$U(1)$ gauge groups, we should have some idea for multi-brane probes. 
One obvious generalization is to replace the $U(1)$ with 
$SU(N)$ gauge groups directly. For the matter content, the
generalization is not so easy. 
A field with charge $(1,-1)$ under gauge groups 
$U(1)_A\times U(1)_B$ and zero for others generalised to a
bifundamental $(N,\bar{N})$ of $SU(N)_A\times SU(N)_B$. However, for
higher charges, e.g., charge 2, we simply do not know 
what should be the generalization in the multi-brane case (for a
discussion on generalised quivers cf. e.g. \cite{Quiver}).
Furthermore, a field with zero charge under all $U(1)$ groups, 
generalises to an adjoint of one $SU(N)$ gauge group in the
multi-brane case, though we do not know which one.

The generalization of the superpotential is also not so
straight-forward. For
example, there is a quartic term in the conifold with nonabelian 
gauge group \cite{Uranga,Greene}, 
but it disappears when we go to the $U(1)$
case. The same phenomenon can happen when treating the generic toric
singularity.

For the examples we give in this paper however, we do not
see any obvious obstruction in the matter contents and superpotential;
they seem to be special enough to be trivially generalized to the
multi-brane case; they are all charge $\pm 1$ under no more than 2
groups. We simply
replace $U(1)$ with $SU(N)$ and $(1,-1)$ fields with bifundamentals
while keeping the superpotential invariant. Generalisations to
multi-brane stack have also been discussed in \cite{Chris}.
\section{Discussions and Prospects}
It is well-known that in the study of the world-volume gauge
theory living on a D-brane probing an orbifold singularity
$\IC^3/\Gamma$,
different choices of decomposition into irreducibles of the space-time
action of $\Gamma$ lead to different matter content and interaction in
the gauge theory and henceforth different moduli spaces (as different
algebraic varieties). 
This strong relation between the decomposition 
and algebraic variety has been shown explicitly for Abelian orbifolds
in \cite{Aspinwall}. It seems that there is only one gauge
theory for each given singularity. 

A chief motivation and purpose of this paper is the realisation that
the above strong statement can not be
generalised to arbitrary (non-orbifold) singularities and in
particular toric singularities.
It is possible that there are several gauge
theories on the D-brane probing the same singularity. 
The moduli space of these inequivalent theories are indeed
by construction the same, as dictated by the geometry of the singularity.

In analogy to the freedom of decomposition into irreps of the group
action in the orbifold case, there too exists a freedom in toric
singularities: any toric diagram is defined only up to a unimodular
transformation (Theorem \ref{iso}). We harness this toric isomorphism
as a tool to create inequivalent gauge theories which live on the
D-brane probe and which, by construction, flow to the same (toric) moduli
space in the IR.

Indeed, these theories constitute another sub-class of examples of
{\em toric duality} as proposed in \cite{toric}. A key point to note
is that unlike the general case of the duality (such as F-D
ambiguities and repetition ambiguities as discussed therein) 
of which we have hitherto little control, these
particular theories are all physical (i.e., guaranteed to be
world-volume theories) by virtue of their being obtainable from the
canonical method of partial resolution of Abelian orbifolds.
We therefore refer to them as {\em phases} of partial resolution.

As a further tool, we have re-examined the Forward and Inverse
Algorithms developed in \cite{Chris,toric,DGM} of extracting the
gauge theory data and toric moduli space data from each other. In
particular we have taken the pains to show what {\em degree of freedom} can
one have at each step of the Algorithm. This will serve to
discriminate whether or not two theories are physically equivalent
given their respective matrices at each step.

Thus equipped, we have re-studied the partial resolutions of the Abelian
orbifold $\IC^3 / (\IZ_3 \times \IZ_3)$, namely the 4 toric del Pezzo
surfaces $dP_{0,1,2,3}$ and the zeroth Hirzebruch surface $F_0$.
We performed all possible $SL(3;\IZ)$ transformation of these toric
diagrams which are up to permutation still embeddable in $\IZ_3 \times
\IZ_3$ and subsequently initiated the Inverse Algorithm therewith.
We found at the end of the day, in addition to the physical theories
for these examples presented in \cite{toric}, an additional one for
both $F_0$ and $dP_2$. Further embedding can of course be done, viz.,
into $\IZ_n \times \IZ_n$ for $n > 3$; it is expected that more phases
would arise for these computationally prohibitive cases, for example
for $dP_3$.

A clear goal awaits us: because for the generic (non-orbifold) 
toric singularity
there is no concrete concept corresponding to the different decomposition of
group action, we do not know at this moment how to classify the phases
of toric duality.
We certainly wish, given a toric singularity, to know (a)
how many inequivalent gauge theory are there and (b) what are 
the corresponding matter contents and superpotential.
It will be a very interesting direction for further
investigation.

Many related questions also arise. For example, by
the AdS/CFT correspondence, we need to understand how to
describe these different gauge theories on the 
supergravity side while the underline geometry is same.
 Furthermore the $dP^2$ theory can be
described in the brane setup by $(p,q)$-5 brane webs
\cite{9710116}, so we want to ask how to
understand these different phases in such brane setups.
Understanding these will help us to get the gauge theory in
higher del Pezzo surface singularities. 

Another very pertinent issue is to clarify the meaning of ``toric
duality.'' So far it is merely an equivalence of moduli spaces of
gauge theories in the IR. It would be very nice if we could make this
statement stronger. For example, could we find the explicit mappings
between gauge invariant operators of various toric-dual theories? 
Indeed, we believe that the study of toric duality and its phase 
structure is worth further pursuit.
\section*{Acknowledgements}
{\it Ad Catharinae Sanctae Alexandriae et Ad Majorem Dei Gloriam...\\}
We would like to extend our sincere gratitute to M. Douglas
for his insights and helpful comments. Also, we would like to thank
A. Iqbal for enlightening discussions. Furthermore, we are thankful to
B. Greene for his comments. And as always, we are
indebted to the CTP and LNS of MIT for their gracious patronage.
In particular we are very grateful to A.~Uranga for pointing out
crucial corrections to the first version of the paper.
\section{Appendix: Gauge Theory Data for $\IZ_n\times \IZ_n$}
For future reference we include here the gauge theory data for the 
$\IZ_n\times \IZ_n$ orbifold, so that, as mentioned in \cite{toric},
any 3-dimensional toric singularity may exist as a partial resolution
thereof.

We have $3n^2$ fields denoted as $X_{ij},Y_{ij},Z_{ij}$ and
choose the decomposition ${\bf 3}\rightarrow (1,0)+(0,1)+(-1,-1)$.
The matter content (and thus the $d$ matrix) 
is well-known from standard brane box
constructions, hence we here focus on
the superpotential \cite{HSU} (and thus the $K$ matrix):
$$
X_{ij}Y_{i(j+1)}Z_{(i+1)(j+1)}-Y_{ij}X_{(i+1)j}Z_{(i+1)(j+1)},
$$
from which the F-terms are
\beq
\label{F-term}
\ba{crcl}
\frac{\partial W}{\partial X_{ij}}: & 
 Y_{i(j+1)}Z_{(i+1)(j+1)} & = & Z_{i(j+1)}Y_{(i-1)j}  \\
&  & & \\
\frac{\partial W}{\partial Y_{ij}}: & 
Z_{(i+1)j} X_{i(j-1)}& =& X_{(i+1)j}Z_{(i+1)(j+1)} \\ & & &  \\
\frac{\partial W}{\partial Z_{(i+1)(j+1)}}: & 
X_{ij}Y_{i(j+1)}& =& Y_{ij}X_{(i+1)j}.
\ea
\eeq

Now let us solve \eref{F-term}. First we have
$Y_{i(j+1)}=Y_{ij}X_{(i+1)j}/X_{ij}$.
Thus if we take $Y_{i0}$ and $X_{ij}$ as the independent 
variables, we have
\beq
\label{solve_Y}
Y_{i(j+1)}=\frac{\prod_{l=0}^{j} X_{(i+1)l}}{\prod_{l=0}^{j} X_{il}}
Y_{i0}.
\eeq
There is of course the periodicity which gives
\beq
\label{constraint_1}
Y_{in}=Y_{i0} \Longrightarrow
 \prod_{l=0}^{n-1} X_{(i+1)l}=\prod_{l=0}^{n-1} X_{il}.
\eeq
Next we use $X_{ij}$ to solve the $Z_{ij}$ as
$Z_{i(j+1)}=Z_{ij} X_{(i-1)(j-1)}/X_{ij},$ whence
\beq
\label{solve_Z}
Z_{i(j+1)}=\frac{\prod_{l=0}^{j} X_{(i-1)(l-1)}}{\prod_{l=0}^{j} X_{il}}
Z_{i0}.
\eeq
As above,
\beq
\label{constraint_2}
Z_{in}=Z_{i0} \Longrightarrow
 \prod_{l=0}^{n-1} X_{(i-1)(l-1)}=\prod_{l=0}^{n-1} X_{il}.
\eeq
Putting the solution of $Y,Z$ into the first equation of \eref{F-term}
we get
$$
\frac{\prod_{l=0}^{j} X_{(i+1)l}}{\prod_{l=0}^{j} X_{il}}
Y_{i0}
\frac{\prod_{l=0}^{j} X_{(i)(l-1)}}{\prod_{l=0}^{j} X_{(i+1)l}}
Z_{(i+1)0}
=
\frac{\prod_{l=0}^{j} X_{(i-1)(l-1)}}{\prod_{l=0}^{j} X_{il}}
Z_{i0}
\frac{\prod_{l=0}^{j-1} X_{il}}{\prod_{l=0}^{j-1} X_{(i-1)l}}
Y_{(i-1)0},
$$
which can be simplified as $Y_{i0} Z_{(i+1)0} X_{i(n-1)}=
Z_{i0} Y_{(i-1)0} X_{(i-1)(n-1)}$,
or $X_{i(n-1)}=X_{(i-1)(n-1)}
\frac{Y_{(i-1)0}}{Y_{i0}}\frac{Z_{i0}}{Z_{(i+1)0}}$.
From this we solve
\beq
\label{X_n-1}
X_{i(n-1)}=X_{0(n-1)} \prod_{l=0}^{i-1}
\frac{Y_{l0}}{Y_{(l+1)0}}\frac{Z_{(l+1)0}}{Z_{(l+2)0}}.
\eeq
The periodicity gives
\beq
\label{constraint_3}
\prod_{l=0}^{n-1}
\frac{Y_{l0}}{Y_{(l+1)0}}\frac{Z_{(l+1)0}}{Z_{(l+2)0}}=1.
\eeq
Now we have the independent variables $Y_{i0}$
$Z_{i0}$ and  $X_{ij}$ for $j\neq n-1$ and $X_{0(n-1)}$,
plus three constraints (\ref{constraint_1})
(\ref{constraint_2}) (\ref{constraint_3}). In fact, considering the
periodic condition for $X$, (\ref{constraint_1}) is equivalent to
(\ref{constraint_2}). Furthermore considering the periodic conditions
for $Z_{i0}$ and $Y_{i0}$, (\ref{constraint_3}) is trivial.
So we have only one constraint. Putting
the expression (\ref{X_n-1}) into (\ref{constraint_1}) we get
$
\prod_{l=0}^{n-2} X_{(i+1)l} 
\frac{Y_{i0}}{Y_{(i+1)0}}\frac{Z_{(i+1)0}}{Z_{(i+2)0}}
=\prod_{l=0}^{n-2} X_{il}
\Rightarrow
\prod_{l=0}^{n-2} X_{(i+1)l}\frac{1}{Y_{(i+1)0}Z_{(i+2)0}}
=\prod_{l=0}^{n-2} X_{il} \frac{1}{Y_{i0}Z_{(i+1)0}}.
$

From this we can solve the $X_{i(n-1)}$ for $i\neq 0$ as
\beq
\label{X_n_2}
X_{i(n-2)}=(\prod_{l=0}^{n-2} X_{0l}) \frac{Y_{i0}Z_{(i+1)0}}{Y_{00}Z_{10}}
(\prod_{l=0}^{n-2} X_{il})^{-1}.
\eeq
The periodic condition does not give new constraints. 

Now we have finished solving the F-term and can summarise the results
into the $K$-matrix. We use the following independent variables:
$Z_{i0}$, $Y_{i0}$ for $i=0,1,...,n-1$; $X_{ij}$ for $i=0,1,...,n-1$
 $j=0,1,...,n-3$ and $X_{0(n-2)}$ $X_{0(n-1)}$, so the total number of
variables is $2n+n(n-2)+2=n^2+2$. This is usually too large to
calculate. For example, even when $n=4$, the $K$ matrix is $48\times
18$. The standard method to
find the dual cone $T$ from $K$ needs to analyse some $48!/(17!31!)$
vectors, which
is computationally prohibitive.
\bibliographystyle{JHEP}

\end{document}